\documentclass[journal]{IEEEtran}
\usepackage{graphicx}
\usepackage{booktabs} 
\usepackage{siunitx} 
\usepackage{amssymb} 
\usepackage[table,xcdraw]{xcolor} 
\usepackage{multirow} 
\usepackage{tikz}
\usepackage{bm}
\usepackage{diagbox} 
\usepackage{makecell} 
\usepackage{amsmath}
\usepackage{pifont}
\usepackage{booktabs}
\usepackage{array} 
\usepackage{multirow} 
\usepackage{graphicx} 
\usepackage[numbers]{natbib}

\ifCLASSINFOpdf
\else
\fi
\begin{document}
%
\title{Modeling Pedestrian Crossing Behavior: A Reinforcement Learning Approach with Sensory Motor Constraints}
%
%
%


\author{Yueyang~Wang,
        Aravinda~Ramakrishnan~Srinivasan,
        Yee~Mun~Lee,
        and~Gustav~Markkula

\thanks{Manuscript received...}
\thanks{Y. Wang is with the Institute for Transport Studies, University of Leeds, Leeds, LS2 9JT UK e-mail: mn20yw2@leeds.ac.uk.}
\thanks{A. Srinivasan, Y. Lee and G. Markkula are with the University of Leeds.}
\thanks{This work was supported by the UK Engineering and Physical Sciences Research Council (grant EP/S005056/1).}}

\maketitle
\begin{abstract}
Understanding pedestrian behavior is crucial for the safe deployment of Autonomous Vehicles (AVs) in urban environments. Traditional pedestrian behavior models often fall into two categories: mechanistic models, which do not generalize well to complex environments, and machine-learned models, which generally overlook sensory-motor constraints influencing human behavior and thus prone to fail in untrained scenarios. We hypothesize that sensory-motor constraints, fundamental to how humans perceive and interact with their surroundings, are essential for realistic simulations. Thus, we introduce a constrained reinforcement learning (RL) model that simulates the crossing decision and locomotion of pedestrians. It was constrained to emulate human sensory mechanisms with noisy visual perception and looming aversion. Additionally, human motor constraint was incorporated through a bio-mechanical model of walking. We gathered data from a human-in-the-loop experiment to understand pedestrian behavior. The findings reveal several phenomena not addressed by existing pedestrian models, regarding how pedestrians adapt their walking speed to the kinematics and behavior of the approaching vehicle. Our model successfully captures these human-like walking speed patterns, enabling us to understand these patterns as a trade-off between time pressure and walking effort. Importantly, the model retains the ability to reproduce various phenomena previously captured by a simpler version of the model. Additionally, phenomena related to external human-machine interfaces and light conditions were also included. Overall, our results not only demonstrate the potential of constrained RL in modeling pedestrian behaviors but also highlight the importance of sensory-motor mechanisms in modeling pedestrian-vehicle interactions.


\end{abstract}

\begin{IEEEkeywords}
Reinforcement learning, Noisy perception, Road user interaction, Pedestrian behavior, Sensory-motor constraints.
\end{IEEEkeywords}


%
\IEEEpeerreviewmaketitle

\section{Introduction}
\label{sec:Intro}

\IEEEPARstart{A}{utonomous} vehicles (AVs) have attracted considerable attention from the public and play a crucial role in developing the intelligent transportation system of tomorrow. A critical aspect of integrating AVs into the urban environment is their ability to interact safely and smoothly with other road users \citep{Rasouli2018AutonomousVT}. Among these road users, pedestrians exhibit complex and often unpredictable behaviors. Thus, for AVs to operate safely in the mixed traffic environment, it is essential to develop models that can help us understand and predict pedestrian behavior \citep{camara2020pedestrian}.

Pedestrian behavior modeling employs primarily two approaches: mechanistic models and machine learning (ML) models. Mechanistic models, including cognitive models grounded in psychology and neuroscience, provide detailed insights into the cognitive processes underlying pedestrian behaviors but often struggle with the complexity and diversity of real-world scenarios \citep{Hollmann2015ACH, markkula2023explaining}. ML models, in contrast, use data-driven techniques to learn from large datasets and predict pedestrian movements effectively. However, they require extensive labeled data, lack interpretability, and sometimes do not generalise well outside their training datasets \citep{Papathanasopoulou2022ADM, quan2021holistic}.

Our previous work has focused on addressing these challenges by developing a pedestrian model that integrates the strengths of both cognitive and ML approaches \cite{Wang2023ModelingHR}. Specifically, we have explored the use of reinforcement learning (RL) to model the binary decision of go/no-go in pedestrian crossing scenarios when interacting with an approaching vehicle, incorporating theory about visual perception to capture realistic human-like road crossing decisions.

In the present work, we aim to further this approach by considering additional aspects of visual perception (external Human-Machine Interface (eHMI) and lighting conditions), and crucially extending it with a biomechanical model of walking, allowing us to address also motor aspects of pedestrian behavior. By integrating these sensory-motor mechanisms into RL models, we attempt to capture a range of behavioral phenomena observed in a controlled experiment on pedestrian road-crossing.

\section{Background}
\label{sec:Background}
\subsection{Pedestrian crossing behaviors}
\label{subsec:Pedestrian crossing behaviors}
\subsubsection{Observed phenomena in pedestrian crossing}
\label{subsubsec:Observed Phenomena in Pedestrian Crossing}

Below, we introduce several key phenomena related to pedestrian crossing as reported by previous literature. Some of these phenomena were modelled in our previous study \cite{Wang2023ModelingHR}, while others remain less explored. One phenomenon we modeled is the gap acceptance rate, which refers to the rate at which pedestrians accept the time or distance gap between themselves and approaching vehicles. Another is the crossing initiation time (CIT), defined as the time between the gap appearing and the pedestrian starting to cross. Both gap acceptance and CIT increase with the time to arrival (TTA) of the vehicle, as well as with the vehicle's speed \citep{tian2022explaining}. Another important metric, pedestrian crossing speed, was not captured by our previous crossing decision model \cite{Wang2023ModelingHR}. This metric tends to decrease as the time gap increases, suggesting a compensatory behavior where pedestrians walk faster when accepting shorter gaps \cite{kalantarov2018pedestrians}.

\subsubsection{Sensory-motor mechanisms in crossing decisions in pedestrian crossing}
\label{subsubsec:Sensory-motor Factors}

Pedestrian crossing decisions are influenced by a variety of sensory-motor mechanisms, which impact how pedestrians perceive and interact with their environment. In this study, we focused on noisy perception, looming aversion, time pressure, walking effort, and ballistic speed control.

\paragraph{Noisy perception}

Human perception of the world is inherently noisy and imperfect \cite{faisal2008noise}. This noisiness can be due to several factors, such as individual differences in sensory acuity \cite{manning2022humans} and environmental conditions (e.g., poor lighting, weather). It has been argued that this visual limitation affected the pedestrian crossing decision \cite{kotseruba2023intend}. For example, noisy perception can lead to errors in estimating the speed and distance of oncoming vehicles, making it challenging for pedestrians to accurately judge safe crossing opportunities. 

\paragraph{Looming aversion}

Visual looming describes the perceived growth of an object's size as it approaches \citep{delucia2008critical}. The aversion to looming refers to the tendency of individuals to react more strongly to objects that appear to be rapidly approaching \cite{mulier2024face}. This phenomenon is rooted in the perceptual system's sensitivity to motion cues that signal potential threats, and influences pedestrian crossing decision. For example, \cite{tian2022explaining} explained speed-dependent gap acceptance behavior by using looming aversion.

\paragraph{Time pressure}

Time pressure has a significant effect on pedestrian crossing behavior \cite{tian2022explaining}. For example, when pedestrians are under time constraints, either due to the rapid approach of a vehicle or a short signal phase \cite{kalantarov2018pedestrians}, they tend to initiate crossing faster and adopt higher walking speeds.

\paragraph{Walking effort}

Individuals choose their walking speed accounting for not only the time but also the energy spent on walking \cite{carlisle2023optimization}. However, the impact of walking effort has not been extensively investigated in the existing pedestrian modeling literature. Considering the trade-off between the energy and time costs can help to better predict the walking dynamics. For example, studies have shown that older pedestrians often exhibit more conservative crossing behaviors, partly due to the increased walking effort required and the need to ensure their safety \cite{oxley2004older}.

\paragraph{Ballistic speed control}

Early researchers noted that humans tend to adjust their sensorimotor movements through intermittent, ballistic control, rather than continuous adjustments \cite{tustin1947nature,craik1948theory}. Specifically, during walking, the legs execute swinging motions that are initiated and completed without active sensory adjustments mid-motion.

\subsection{Pedestrian crossing models}
\label{subsec:Pedestrian crossing models}

The development of models to predict and understand pedestrian crossing behavior has been a significant focus of research for many years. These models vary in complexity and approaches, capturing different aspects of pedestrian decision-making processes and walking dynamics. 

\subsubsection{Logistic regression models} Logistic regression models, including more recent cognitive cue-based models like the one proposed in \cite{tian2022explaining}, aim to predict whether a pedestrian will cross the street under certain conditions. These models typically consider factors such as vehicle speed, and distance \cite{lobjois2007age}. However, these models primarily focus on crossing decisions without considering the subsequent walking dynamics, making them difficult to integrate into more comprehensive simulations of traffic and pedestrian behavior.

\subsubsection{Mechanistic models} More complex mechanistic models, such as evidence accumulation models, attempt to simulate the human decision-making process by considering how pedestrians accumulate information over time to make crossing decisions \cite{giles2019zebra,pekkanen2022variable}. These models provide a deeper understanding of the cognitive processes involved, and in some cases also include some limited consideration of walking dynamics  \cite{markkula2023explaining}. However, integrating multiple cognitive theories into a unified model is challenging because these theories often have differing assumptions, involve complex interactions, require scalable and adaptable frameworks, and increase model complexity, which can reduce practical usability. This complexity makes it difficult to capture a broader range of behaviors and apply such models to more complex situations and environments \cite{markkula2023explaining}.

\subsubsection{Machine Learning Models} ML models leverage large datasets and advanced algorithms to predict pedestrian behavior with high accuracy and have shown good performance in real-time pedestrian trajectory prediction \cite{quan2021holistic}. While the accuracy of these ML models can be impressive, they share several limitations. First, they often act as 'black boxes', meaning they may fail to reveal the reasons or mechanisms behind those behaviors \cite{Madala2023MetricsFM}. Second, high-level statistical accuracy does not guarantee that the models capture those behaviors that matter to humans \citep{srinivasan2023beyond}. Third, the absence of a theoretical grounding of the behavior can sometimes lead to performance issues. While they perform well on the training data, they may struggle to generalise to new and unseen data \citep{xu2020explainable}. In addition, ML models heavily rely on large training datasets. Whereas collecting such extensive data under all possible road conditions is almost an impossible task, leading to critical scenarios being underrepresented or missing in most datasets \citep{diaz2022ithaca365}.

\subsubsection{Models based on (bounded) optimality}

Another approach to modeling human behavior is \emph{bounded optimality} \cite{markkula2023explaining,hoogendoorn2003simulation, wang2024pedestrian}, based on the assumption that humans behave optimally with respect to a utility or cost function, but with constraints imposed by the human cognition and body \citep{gershman2015computational,oulasvirta2022computational}. By using RL as a method for solving the bounded optimality problem, this approach integrates the strengths of both cognitive models and ML models.

Different from traditional data-driven ML algorithms, which typically learn directly from large datasets without iterative interaction with the environment, RL offers a paradigm wherein an agent interacts with a dynamic environment, and the optimal policy will be derived through trial-and-error~\citep {kaelbling1996reinforcement}. RL is particularly used for modeling tasks that involve sequential decision-making, making it inherently suitable for tasks like the pedestrian crossing, which require a series of decisions based on changing environmental conditions, especially if suitably constrained by models of human sensory and motor mechanisms.

Our previous work using this method captured the pedestrian crossing decisions when interacting with one vehicle, and captured several observed behavioral phenomena \cite{wang2024pedestrian}. However, research gaps remain in the previous model. Key questions include whether our integrated cognitive and RL approach can generalise to a wider range of situations and phenomena, and whether the model can be expanded to include motor constraints, which would allow us to more accurately represent the physical execution of walking actions during road crossing.

\section{Methods}
\label{sec:Method}

\subsection{Experiment}
\label{subsec:Experiment}

To develop and test our model, we used a dataset from an experiment conducted in the University of Leeds Highly Immersive Kinematic Experimental Research (HIKER) laboratory.


\begin{figure*}
    \vspace{-0.4cm} 
      \centering
      \includegraphics[scale=0.5]{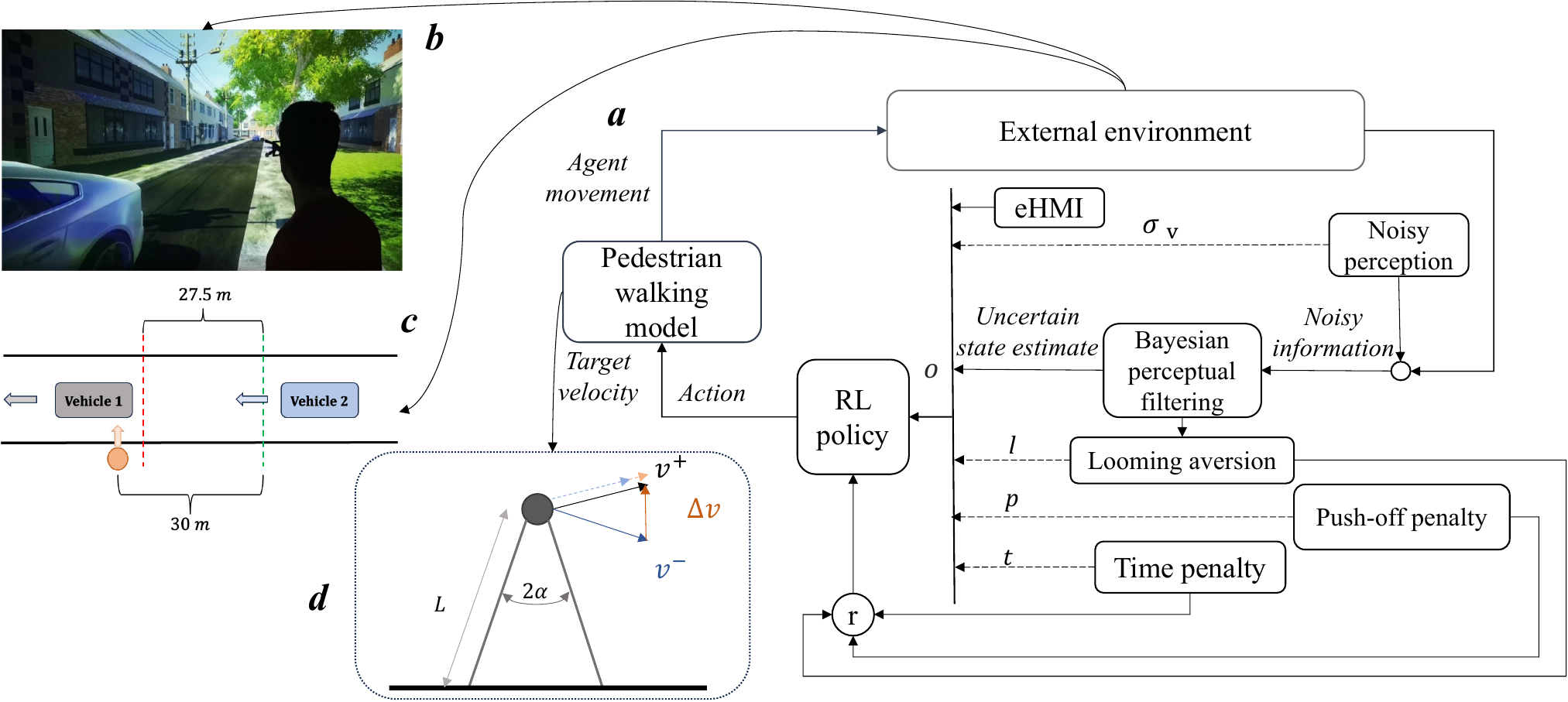}
      \vspace{-0.2cm}
      \caption{Panel (a) is the Model framework. Panel (b)  is the virtual environment. Panel (c) is a schematic of the deceleration procedure used in this study. Panel (d) is the walking model, adopted from \cite{carlisle2023optimization}}
      \label{fig:Model framework}
      \vspace{-0.4cm}
\end{figure*}

The experiment is described in full in \cite{lee2024Under}, below we provide a summary for completeness. In this study, the task of the participant was to cross the road between two approaching vehicles safely – as shown in panel (b) of \figurename\ref{fig:Model framework}. A mixed design was used, with five within-participant variables: (i) the initial speed $v_0$ of the approaching vehicles (25/30 mph); (ii) the time gap $t_0$ between the vehicles (3/5 s); (iii) the yielding behavior of the second vehicle with a constant deceleration rate $a$ (yielding/non-yielding); (iv) the presence of eHMI while yielding (present/absent); and (v) the time of day (daytime/nighttime); and one between-participant variable: participants’ age (younger/older) \cite{lee2024Under}. In this study, the eHMI was presented as a cyan color light around the windscreen in some of the yielding trials. Participants were informed that the presence of eHMI means that the approaching vehicle was signalling \emph{‘I am yielding’.} Prior to the experiment, all participants signed the consent form agreeing to take part in the study. Ethical approval was obtained from the University of Leeds Research Committee.

The recorded data used from the experiment include participants' crossing decisions, CIT, and walking speeds; these quantities were all estimated from a body tracker located on the participant's head, measuring XYZ position at a sampling rate of 50 Hz. Furthermore, we classified crossing behaviors in yielding scenarios as either early or late. This classification relies on the concept of bimodal crossing, where pedestrian responses can typically be categorised into two distinct modes based on their timing relative to vehicle behaviors, with pedestrians choosing to cross either well before a vehicle slows down or waiting until it is clearly safe to do so \cite{giles2019zebra}. In our experiment, the threshold for this behavior was quantified by whether the pedestrian began crossing before the vehicle decelerated to two-thirds of its initial speed, which cleanly separate these two crossing modes in our data. For a more detailed explanation of our methodology, please refer to \cite{lee2024Under}.


\begin{table}[b]
    \vspace{-0.4cm}
    \centering
    \caption{Vehicle approach scenarios in the experiment.}
    
    \begin{tabular}{cccc}
        \toprule
        \textbf{Scenario type} & \bm{$v_0~(mph)$} & \bm{$\tau_0~(s)$} & \bm{$a~(m/s^2)$} \\
        \midrule
        Constant speed  & $25$  & $3$ & $N/A$ \\
        $ $  & $25$ & $5$ & $N/A$ \\
        $ $  & $30$  & $3$ & $N/A$ \\
        $ $  & $30$ & $5$ & $N/A$ \\
        Yielding  & $25$  & $3$ & {-$2.3$}\\
        $ $  & $25$ & $5$ & {-$2.3$}\\
        $ $  & $30$  & $3$ & {-$2.3$}\\
        $ $  & $30$ & $5$ & {-$2.3$}\\

        \bottomrule
    \end{tabular}       
    \vspace{-0.4cm}
    \label{tab:scenarios}
\end{table}

\subsection{Model}
\label{subsec:Model}


\subsubsection{Sensory-motor mechanisms}
\label{subsubsec:Sensory-motor factors}
In this paper, we explore the influence of four sensory-motor mechanisms on the crossing behavior as detailed in Section \ref{subsubsec:Sensory-motor Factors}, noisy perception, visual looming, time pressure, walking effort and ballistic speed control; as illustrated in \figurename~\ref{fig:Model framework}) (a).

\paragraph{Noisy perception} The agent's perception of environmental distances includes a constant Gaussian angular noise, $\sigma_\mathrm{v}$, which affects the estimation of vehicle positions and velocities. This model incorporates a Kalman filter to adaptively refine these estimates, leveraging Bayesian methods to approximate the state of the environment refer to the previous paper for mathematical details \cite{wang2024pedestrian}.

\paragraph{Visual looming}This phenomenon is mathematically represented as inverse $\tau$—the ratio of a vehicle’s optical expansion rate to its size on the observer's retina, which is an estimate of the inverse TTA \citep{markkula2016farewell,delucia_2015}. We defined looming aversion as follows: \(L = c \cdot \frac{1}{\hat{\tau}}\), where \( c \) is the weight of the looming aversion, if  \( v>0 \); otherwise 0, and $\hat{t}$ Kalman-filter estimated TTA of the vehicle.

\paragraph{Time pressure} The time pressure in our model is calculated as: \(T = \alpha \cdot t\), where \( t \) is the time elapsed in the episode and \( \alpha \) is the scaling factor of the time pressure.

\paragraph{Walking effort} To account for walking effort, we adopt the biomechanical model of walking from \cite{carlisle2023optimization}. The equation for the new velocity $v_i^{+}$ is: 
\begin{equation}
\label{eq: new vel}
v_i^{+} = v_i^{-} \cos 2\alpha + \sqrt{2u_i} \sin 2\alpha
\end{equation}
where \( v_i^{-} \) is the initial walking speed, \( v_i^{+} \) is the new walking speed after the action, and 2\( \alpha \) is the angle between two legs. The term \(\sqrt{2u_i}\) represents the velocity component due to the exerted effort. The derivation of Equation \ref{eq: new vel} starts with the assumption that the pedestrian's leg motion can be described as a pendulum (see panel (d) of \figurename\ref{fig:Model framework}), and that the speed change is caused by the effort \( u_i \). This effort required for this change is related to the kinetic energy: 
\begin{equation}
\label{eq: kinetic energy 1}
u_i = \frac{1}{2} m (\Delta v)^2
\end{equation}
and solving for \( \Delta v \):
\( \Delta v = \sqrt{2u_i / m} \). Since \( u_i \) is effort per unit mass, this simplifies to:
\( \Delta v = \sqrt{2u_i} \). Therefore, the effort \( u_i \) required for this change is given by:

\begin{equation}
\label{eq: kinetic energy 2}
u_i = \frac{{(v^{-} \cos(2\alpha) - v^{+})}^2}{2 \cdot \sin^2(2\alpha)}
\end{equation}
The total walking effort, \( E_w \), is then calculated as: \(E = \beta \cdot u_i\), where \( \beta \) is a parameter that scales the walking effort for different individuals.

\paragraph{Ballistic speed control}
As mentioned in Section \ref{subsubsec:Sensory-motor Factors}, we assumed that the agent adjust their movement using the ballistic speed control. Therefore, upon selecting an action, the agent employs ballistic speed control, which means the agent maintains the acceleration rate $a$ needed to change its speed to the desired value, which is defined as \( a = \frac{V_t - V_{t-1}}{t} \), where $V_t$ and $V_{t-1}$ represent the velocities of the agent at two consecutive decision points. The time interval $t$ is calculated based on the relation between step length, velocity, and frequency, with the preferred speed \( v \) and step length \( s \) following the equation \( s = v^{0.42} \) \cite{grieve1968gait}.


\subsubsection{Reinforcement learning problem}
\label{subsubsec: Reinforcement learning problem}
Our RL environment is an example of a Partially Observable Markov Decision Process (POMDP), where the agent does not have direct access to the true state; rather, it receives observations that may only partially or noisily reflect the actual state. The POMDP is represented by a tuple $<S,A,T,R,O>$, where $S$ is a set of states, $A$ is a set of actions, $T$ is the transition function, $R$ is the reward function and $O$ is the set of possible observations received by the agent.

\paragraph{State}
At each time step $t$, the environment is in a state $s_t \in S$, which includes the position and velocity of both the pedestrian (agent) and two vehicles. All model variants in our study share the same state space: the pedestrian's position \(x_\mathrm{p}, y_\mathrm{p}\), the vehicle's positions \(x_{\text{veh}_1},x_{\text{veh}_2}, y_{\text{veh}_1}, y_{\text{veh}_2}\), the vehicle's velocities \(v_1, v_2\), and time step $t$. The simulation updates the state every 0.1 s, a step size chosen as a tradeoff between the computational cost and the accuracy to represent dynamic interactions between vehicles and pedestrians.

\paragraph{Action}
At the completion of each walking step \( t \), the agent executes an action \( a_t \) from set \( A \). In this study, the action space in this study is \( A = \{0.1, 0.2, \ldots, 2 \} \) m/s. Each action value corresponds to a different desired walking speed, enabling the agent to adjust its velocity. Upon selecting an action, the agent employs ballistic speed control, which means the agent maintains the acceleration rate needed to change its speed to the desired value, which is defined as \( a = \frac{V_t - V_{t-1}}{t} \). After the decision is made, the simulation progresses until the execution of the walking step is completed, at which point the agent can make the next decision.

\paragraph{Transition}
The transition function defines how the current state \( s_t \) changes to the next state \( s_{t+1} \) based on the action \( a_t \). The vehicle's movement follows kinematic equations with the speeds and accelerations corresponding to the scenario in question. The agent changes its speed linearly to the target speed, as mentioned in the last paragraph. The simulation ends when a collision occurs or the agent crosses the road safely. 

\paragraph{Reward}
According to the experiment, we assumed that the agent wanted to cross safely, while minimizing time losses, energy losses, and any discomfort from experiencing high levels of visual looming.

The reward function $r$ was designed accordingly: \(r = A - C - E - L\), where A is 20-T if arrival, otherwise 0; C is 20 if collision, otherwise 0;  \(L = c \cdot \frac{1}{\hat{\tau}}\) if  \( v>0 \), otherwise 0,  and \(E = \beta \cdot u_i\), where \( E \) is the penalty for walking effort. The reward \( r \) is bounded within the range \([-20, +20]\) to prevent extreme values from affecting the model training.

\paragraph{Observation}
As previously mentioned, the agent observes own state and Kalman estimates of the vehicle's state. Moreover, because we have trials with eHMI on the vehicle, we also gave the input representing eHMI in the model, with 0 for eHMI off and 1 for eHMI on. In addition, since we have different parameters that influence the sensory-motor process, i.e., $\sigma_\mathrm{v}$, \(\alpha \), \(\beta \) and \( c \), we also gave these parameters as input to the RL policy, i.e., we are conditioning the RL on these parameters \cite{howes2023towards,Li2023Modeling}, which we refer to as non-policy parameters to distinguish them from the parameters of the policy neural network (connection weights and biases).

\subsubsection{Different Model Variants}
\label{subsubsec: Different Model Variants}

To explore how the various mechanistic assumptions in the model affect the generated pedestrian behavior, we tested different model variants:

\emph{SM:} A model with all of the mechanistic assumptions described above, including both sensory assumptions (visual limitations and looming aversion) and motor assumptions (walking effort and ballistic speed control)

\emph{S:} A model with the sensory assumptions but not the motor assumptions. When excluding the assumption about ballistic speed control, a more conventional RL approach of directly controlling the speed at each time step \cite{vizzari2022pedestrian,Sun2019Crowd,xu2020local} was used.

\emph{M:} A model with the motor assumptions but not the sensory assumptions.

\subsubsection{Reinforcement learning}
\label{subsubsec: Reinforcement learning network}
Proximal Policy Optimization (PPO) is a policy gradient RL algorithm that improves training stability and reliability by using a clipped objective function \cite{schulman2017proximal}. This method strikes a balance between exploration and exploitation, ensuring that updates to the policy are not too large. PPO is computationally efficient and straightforward to implement, making it a popular choice for various reinforcement learning tasks. 

For our implementation, we used the Stable Baselines 3 (SB3) library \cite{stable-baselines3}. The model was trained for 3 million environment time steps using the PPO algorithm. We used a network architecture with 128 neurons in the first hidden layer and 64 neurons in the second hidden layer. All other hyperparameters were kept at their SB3 PPO default values. As mentioned, we gave non-policy parameters as additional input to the RL network; for each new RL episode we sampled uniformly from these ranges: [$\sigma_\mathrm{v}$ (0-10), \(\alpha \) (0-4), \(\beta \) (0-10), and \( c \) (0-10)]. In practice, this means that the RL is learning how the optimal policy varies across this space of non-policy parameters.

To ensure the agent did not learn trivial policies based on limited experimental conditions, we trained the RL agent with a wider range of kinematic condition than the scenarios in the experiment. The initial speed was sampled from a uniform distribution between 8 m/s and 17 m/s, and the time gap was sampled from a uniform distribution between 0.1 s and 10 s.

\subsubsection{Fitting non-policy parameters}
\label{subsubsec: Fitting of the non-policy parameters}

The correct values for the non-policy parameters are unknown and may vary among individual participants in the experiment. Additionally, considering the day and night scenarios in the experiment, we hypothesised that visual ability differs between these scenarios. Consequently, for each participant, we wished to fit two $\sigma_\mathrm{v}$ values (one for day and one for night) and one value for each of the other three parameters.

To fit these parameters to data, we used Bayesian Optimisation for Likelihood-Free Inference (BOLFI), a method employed for parameter estimation in models where the likelihood function is intractable or computationally expensive to evaluate \cite{gutmann2016bayesian}. This inference method has been previously applied to RL-based simulation models of human behavior \cite{kangasraasio2017inferring,moon2022speeding}. BOLFI requires the user to define a function quantifying the discrepancy between simulated and observed data, and employs a Gaussian process as a surrogate model to approximate how this discrepancy function varies with model parameter values.

In our study, we used BOLFI to find non-policy parameter values minimising the following discrepancy function:

\begin{equation}
\label{eq:Discrepancy}
\begin{split}
\text{Discrepancy} = \ \log \Bigg( &\left( \sum_{i=1}^{n_{\text{ny}}} \frac{\left| \text{g}_i - \hat{\text{g}}_i \right|}{\text{g}_{\max}} \right)
+ \left( \sum_{i=1}^{n_y} \frac{\left| \text{e}_i - \hat{\text{e}}_i \right|}{\text{e}_{\max}} \right) \\
& \!\!\!\!\!\!\!\!\!\!\!\!\!\!\!\!\!\!\!\!\!\!\!\!\!\!\!\!\!\!\!\!\!\!\!\!\!+ \left( \sum_{i=1}^{n} \frac{\left| \text{CIT}_i - \hat{\text{CIT}}_i \right|}{\text{CIT}_{\max}} \right) 
+ \left( \sum_{i=1}^{n} \frac{\left| \text{v}_i - \hat{\text{v}}_i \right|}{\text{v}_{\max}} \right) \Bigg)
\end{split}
\end{equation}

where the \( \hat{} \) refer to model predictions, \emph{n}, \emph{n\textsubscript{ny}}, \emph{n\textsubscript{y}} are the number of all trials, non-yielding trials, and yielding trials for one participant.

Each term in Equation \ref{eq:Discrepancy} represents the difference between the average values of specific behavioral metrics under different kinematic conditions, as generated by various combinations of non-policy parameter values in the simulation and as observed in the human data, which our model aims to reproduce. These metrics include the gap acceptance rate (the frequency of crossing before the second vehicle in constant-speed scenarios), early crossing rate (the frequency of crossing early vs late in yielding scenarios, CIT, and average walking speed. Dividing by the maximum observed value normalises the differences, as each metric can have a different value range.

We initialised BOLFI with uniform distributions for all the non-policy parameters, in the ranges mentioned in Section \ref{subsubsec: Reinforcement learning network}, and ran 80 optimisation iterations, which was sufficient to achieve convergence for all model variants.

\subsubsection{behavioral phenomena}
\label{subsubsec: behavioral phenomena}
For the analysis of the behavioral phenomena both in the experiment and model, we used Generalized Linear Mixed Models (GLMM) for binary dependent variables (gap acceptance and early crossing rate) and Linear Mixed Models (LMM) for continuous dependent variables (CIT and average walking speed). The fixed effects variables included \emph{time of day (day / night)}, \emph{Time Gap (3 s / 5 s)}, \emph{Speed (25 mph / 30 mph)}, and \emph{eHMI (on / off)} where applicable. Participant ID was included as a random effect variable in all models. As we observed different pedestrian behaviors in early and late crossing in yielding scenarios, we used early or late crossing as an additional fixed effect in the analysis of average crossing speed, coding 0 for late crossing and 1 for early crossing.

\section{Results}
\label{sec: Result}

\begin{table}[b]
\vspace{-0.4cm}
\centering
\caption{Gap acceptance rate in non-yielding scenarios and early crossing rate in yielding scenarios. (* \( p < 0.05 \), ** \( 0.01 > p > 0.001 \), *** \( p < 0.001 \)).}
\label{tab:gap acceptance and early crossing}
\resizebox{\columnwidth}{!}{%
\begin{tabular}{lcccccccc}
\toprule
 & \multicolumn{4}{c}{Gap acceptance} & \multicolumn{4}{c}{Early Crossing} \\
\cmidrule(lr){2-5} \cmidrule(lr){6-9}
 & Estimate & Std. Error & p-value &  & Estimate & Std. Error & p-value \\
\midrule
Intercept   & -19.409 & 1.965 & {$< .001$}  &  *** & 17.817 & 1.627 & {$< .001$} & *** \\
Speed       & 0.371 & 0.09926 & {$< .001$} & *** & 0.426  & 0.090 & {$< .001$} & *** \\
Time Gap    & 3.089 & 0.244 & {$< .001$} & *** & 2.576  & 0.173 & {$< .001$} & *** \\
Time of Day & -0.45262 & 0.217 & 0.037  &  * & -0.410 & 0.194 & 0.035 & * \\
eHMI        & /      & /        & /      &   & -0.330 & 0.194 & 0.0885 &  \\
\bottomrule
\end{tabular}%
}
\vspace{-0.4cm}
\end{table}

\begin{table*}
\centering
\caption{Crossing Initiation Time of different conditions}
\label{tab:crossing_initiation_time_decel}
\begin{tabular}{lcccccccccccc}
\toprule
 & \multicolumn{4}{c}{Non-yielding} &\multicolumn{4}{c}{Early Crossing} & \multicolumn{4}{c}{Late Crossing} \\
\cmidrule(lr){2-5} \cmidrule(lr){6-9}\cmidrule(lr){10-13}
 & {Estimate} & {Std. Error} & {p-value} &  &{Estimate} & {Std. Error} & {p-value} &  & {Estimate} & {Std. Error} & {p-value} \\
\midrule
Intercept  & -0.439 & 0.099 &{$< .001$}  & ***  & -1.514 & 0.402 & {$< .001$} & *** & 1.435 & 0.425 & {$< .001$} & ***\\
Speed    & 0.042 & 0.006 & {$< .001$} & ***   & 0.099 & 0.028 & {$< .001$} & ***& 0.094 & 0.030 & 0.001  & **   \\
Time Gap  & 0.105 & 0.009 & {$< .001$}& ***   & 0.197 & 0.041 & {$< .001$}& *** & 1.006  & 0.037 & {$< .001$} & *** \\
Time of Day  & -0.062 & 0.014 & {$< .001$}& ***    & -0.151 & 0.061 & 0.014  &  * & 0.193 & 0.066 & 0.003 & * \\
eHMI      &   &  &     &     & -0.105 & 0.061 & 0.087   &   & -0.581 & 0.066 & {$< .001$} & *** \\
\bottomrule
\end{tabular}
\vspace{-0.4cm}
\end{table*}

\subsection{Experimental results}
\label{subsec: Experimental results}

In this section, we present the experimental findings for the four behavioral metrics introduced in Section \ref{subsubsec: Fitting of the non-policy parameters}: gap acceptance rate in non-yielding scenarios, early crossing rate in yielding scenarios, CIT and average walking speed. We only report statistically significant effects (\( p < 0.05 \)) in this section.

\begin{table}[!b]
\centering
\caption{Average crossing speed of different conditions}
\label{tab: Average speed of different conditions}
\begin{tabular}{lcccc}
\toprule
 & {Estimate} & {Std. Error} & {p-value} & \\
\midrule
Intercept  & 1.359 & 0.037 & {$< .001$}  & ***  \\
Speed    & -0.001 & 0.002 & 0.609 &    \\
Time Gap  & -0.046 & 0.003 & {$< .001$} & ***  \\
Time of Day  & -0.004 & 0.005 & 0.447  &   \\
Early cross  & 0.188 & 0.008 & {$< .001$} & ***  \\
eHMI      & -0.013 & 0.005 & 0.016  &   *  \\
\bottomrule
\end{tabular}
\vspace{-0.4cm}
\end{table}

\subsubsection{Gap acceptance and early crossing rate}

As can be seen in the first row of \figurename~\ref{fig: Gap_Early_Combined}, the gap acceptance rate and early crossing rate showed the same pattern: Pedestrians were more likely to cross when approaching vehicles were travelling at 30 mph compared to 25 mph (\( p < .001 \)), and to cross with a larger time gap (\( p < .001 \)). These trends are consistent with previous studies mentioned in Section \ref{sec:Background}. Additionally, pedestrians were more inclined to cross during the daytime compared to nighttime (\( p < 0.05 \)).

\subsubsection{Crossing initiation time (CIT)}

As can be seen in the first row of \figurename~\ref{fig: CIT_Combined}, pedestrians were more likely to cross earlier when the speed of approaching vehicles was higher (\( p < .001 \)), and when the time gap was greater (\( p < .001 \)).  These trends were consistent across the three categories. In addition, pedestrians showed a smaller CIT at nighttime compared to daytime in non-yielding and early crossing in yielding scenarios. Conversely, in late crossing in yielding scenarios, pedestrians initiated crossing more slowly at nighttime compared to daytime. Furthermore, from \tablename~\ref{tab:crossing_initiation_time_decel}, eHMI had an effect on CIT in late crossing scenarios (\( p < .001 \)). Specifically, when eHMI was on, pedestrians started to cross earlier.

\subsubsection{Average walking speed}

From the first row of \figurename~\ref{fig: Spd_Combined}, it can be seen that pedestrians crossed the road faster when the time gap was smaller (\( p < .001 \)). Pedestrians were also more likely to cross faster during early crossings (\( p < .001 \)). Furthermore, the average crossing speed was also affected by the eHMI (\( p < .001 \)). Specifically, when eHMI was on, pedestrians walked slower.

Overall, the experimental data showed 19 different statistically significant effects, shown as cells shaded gray in \figurename~\ref{fig: Summary_table}.

\subsection{Model results}

\begin{figure*}
      \centering
      \includegraphics[scale=0.3]{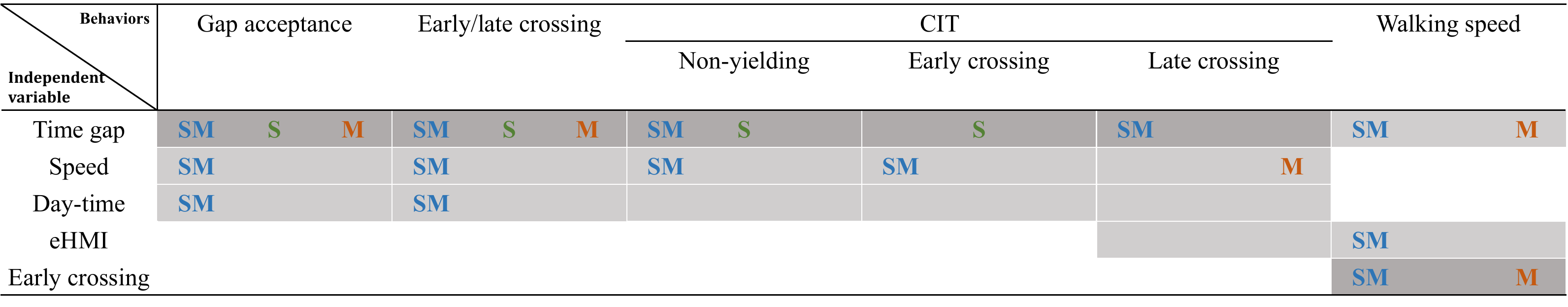}
      \vspace{-0.2cm}
      \caption{Key behaviors in the experiment. The gray cells indicate the statistically significant effects of the independent variable on pedestrian behaviors observed in the experiment, with darker gray denoting a larger effect size (greater than 0.40) and lighter gray representing a smaller effect size (smaller than 0.40). Different color symbols represent different model variants. The cell with the gray color and symbols means that the corresponding model variant captures that phenomenon.}
      \label{fig: Summary_table}
      \vspace{-0.4cm}
\end{figure*}

In this section we examine to what extent the three different model variants were able to capture these observed effects; this is also summarised in \figurename~\ref{fig: Summary_table}.

\subsubsection{Gap acceptance and early crossing rate}

\begin{figure}[!b]
    \vspace{-0.4cm} 
      \centering
      \includegraphics[scale=0.3]{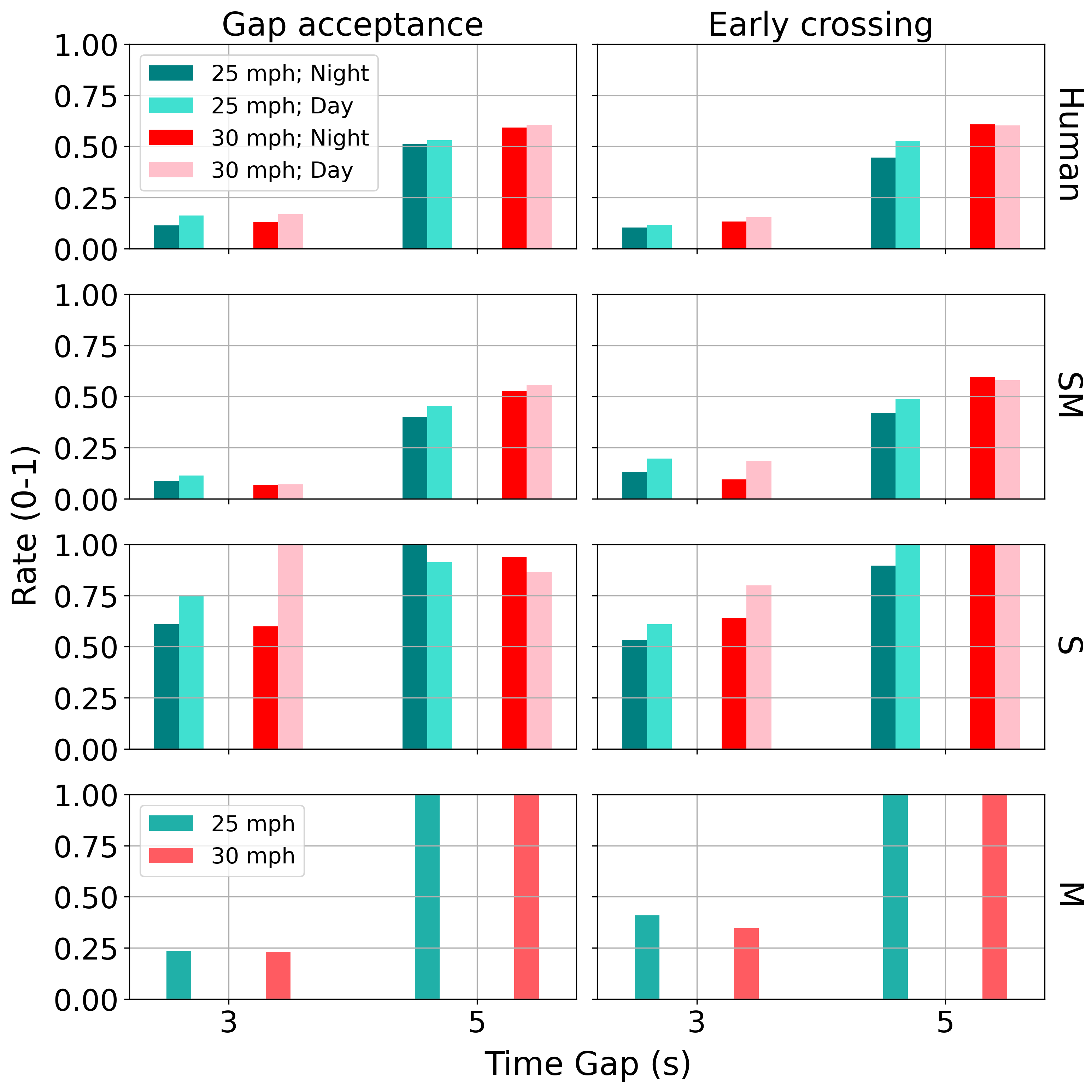}
      \vspace{-0.2cm}
      \caption{Gap acceptance rate in non-yielding scenarios and early crossing rate in yielding scenarios as observed in human behavior and predicted by the models.}
      \label{fig: Gap_Early_Combined}
      \vspace{-0.4cm}
\end{figure}

Rows 2-4 of \figurename~\ref{fig: Gap_Early_Combined} show the gap acceptance behavior of the three model variants: The experimental results for gap acceptance and early crossing rates were best captured by the SM model. This model captured all six phenomena. In contrast, the S and M models only captured two of these phenomena: the increases with increasing time gaps, but not the speed or day/night effects.

\begin{figure*}
      \centering
      \includegraphics[scale=0.25]{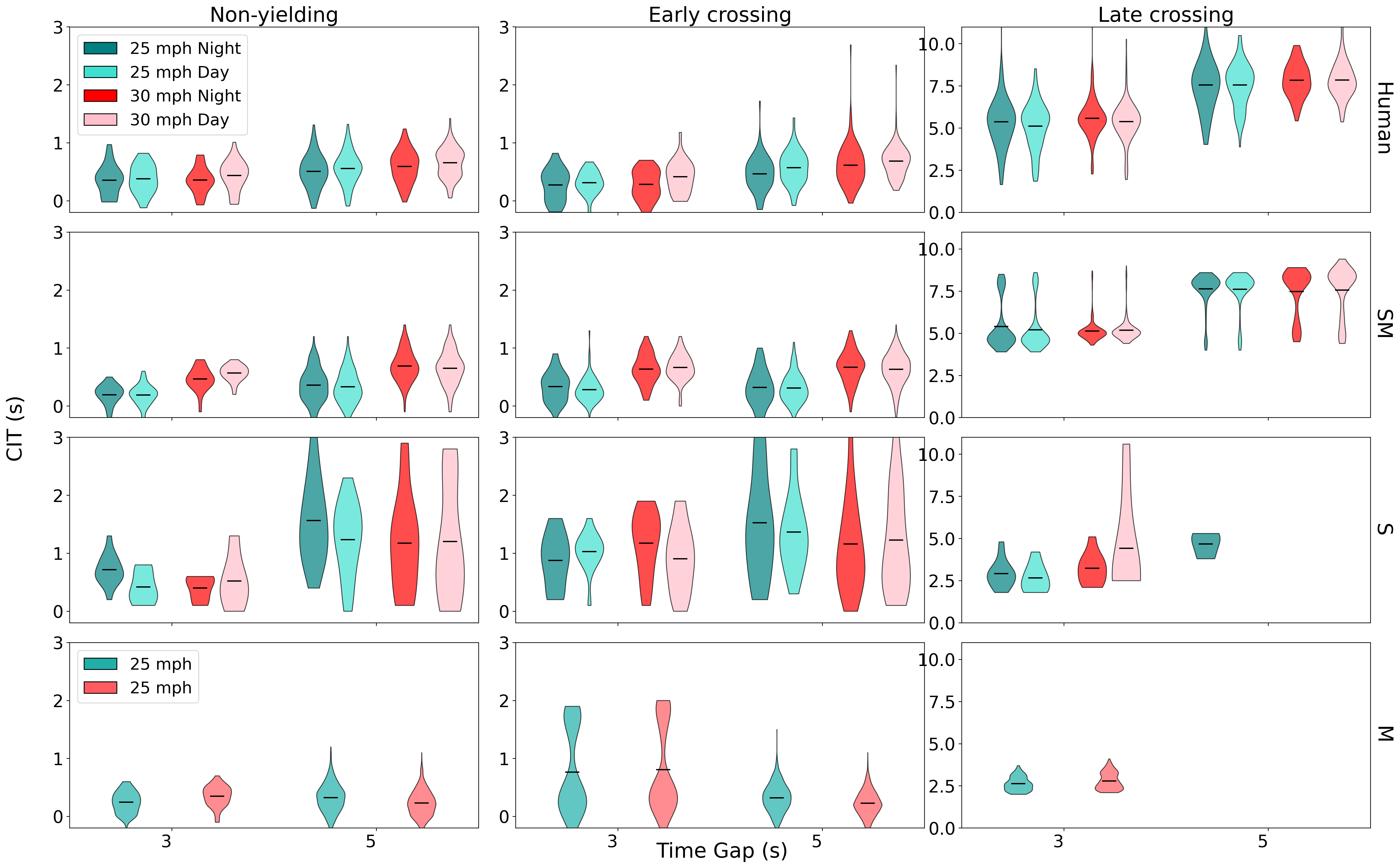}
      \caption{Crossing Initiation Time (CIT) under different conditions. This figure shows the distribution of CIT under different conditions. The first row shows human data, the second row presents the VLW model result (accounting for walking effort), and the third row shows the VL model result (without walking effort). The blue lines within each violin plot represent the mean values. Notably, in the late crossing category for the VL model, no CIT data is available for a 5-second time gap at 30 mph, indicating that no agents chose to cross late under this condition.}
      \label{fig: CIT_Combined}
      \vspace{-0.4cm}
\end{figure*}

\begin{figure*}
    \vspace{-0.4cm} 
      \centering
      \includegraphics[scale=0.25]{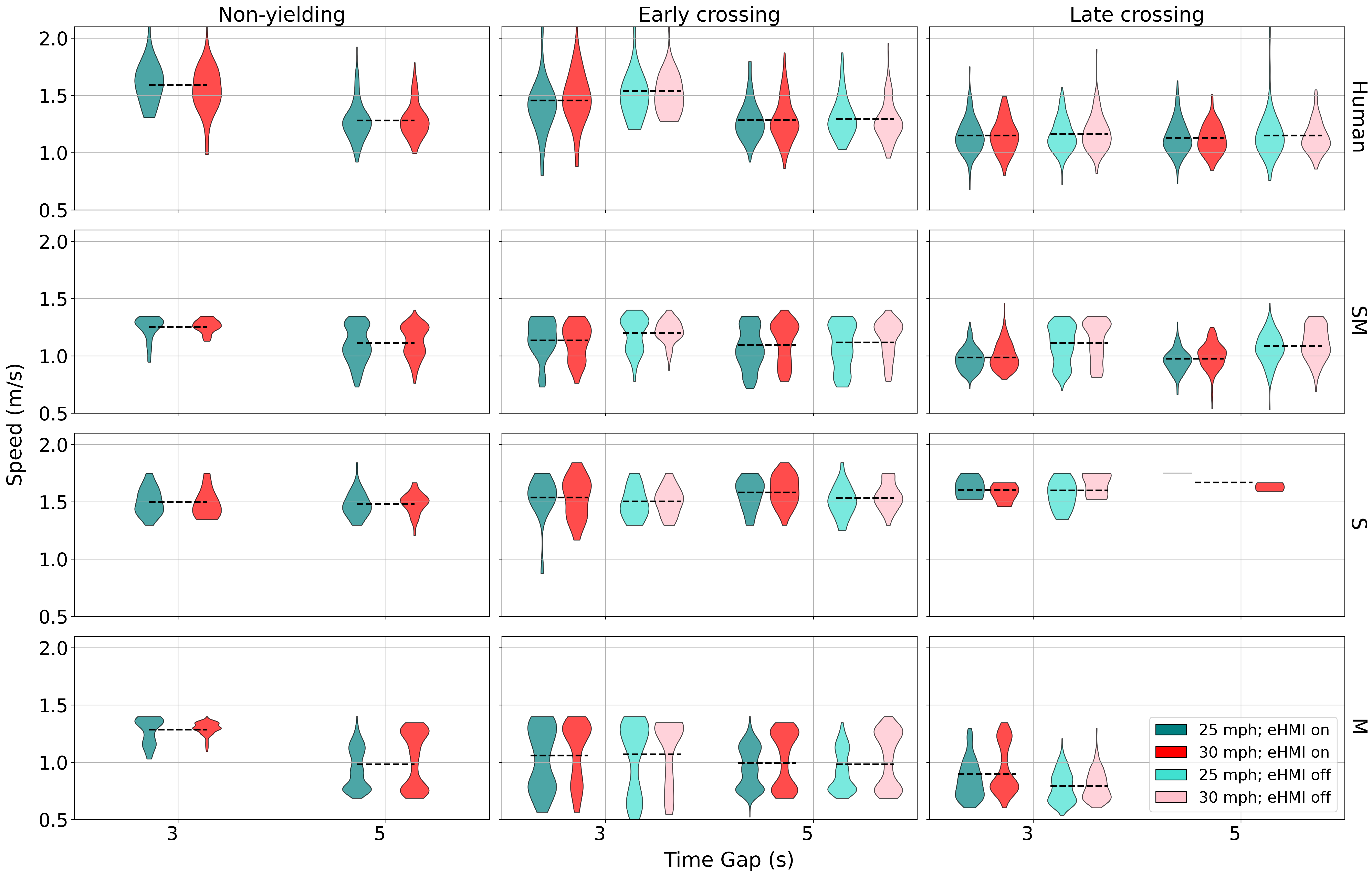}
      \vspace{-0.2cm}
      \caption{Average walking speed. This figure illustrates the average walking speeds for different crossing decisions—non-yielding, early crossing, and late crossing—under different vehicle speeds (25 mph and 30 mph) and eHMI conditions (on and off). Each pair of violins represents different vehicle speeds (25 mph and 30 mph) within the same time gap and eHMI on/off condition. The black dashed lines indicate the mean walking speed for each pair of violins. This pairing method is used because vehicle speed did not have a significant effect on walking speed in the experiment.}
      \label{fig: Spd_Combined}
      \vspace{-0.4cm}
\end{figure*}

\begin{figure}
    \vspace{-0.4cm} 
      \centering
      \includegraphics[scale=0.25]{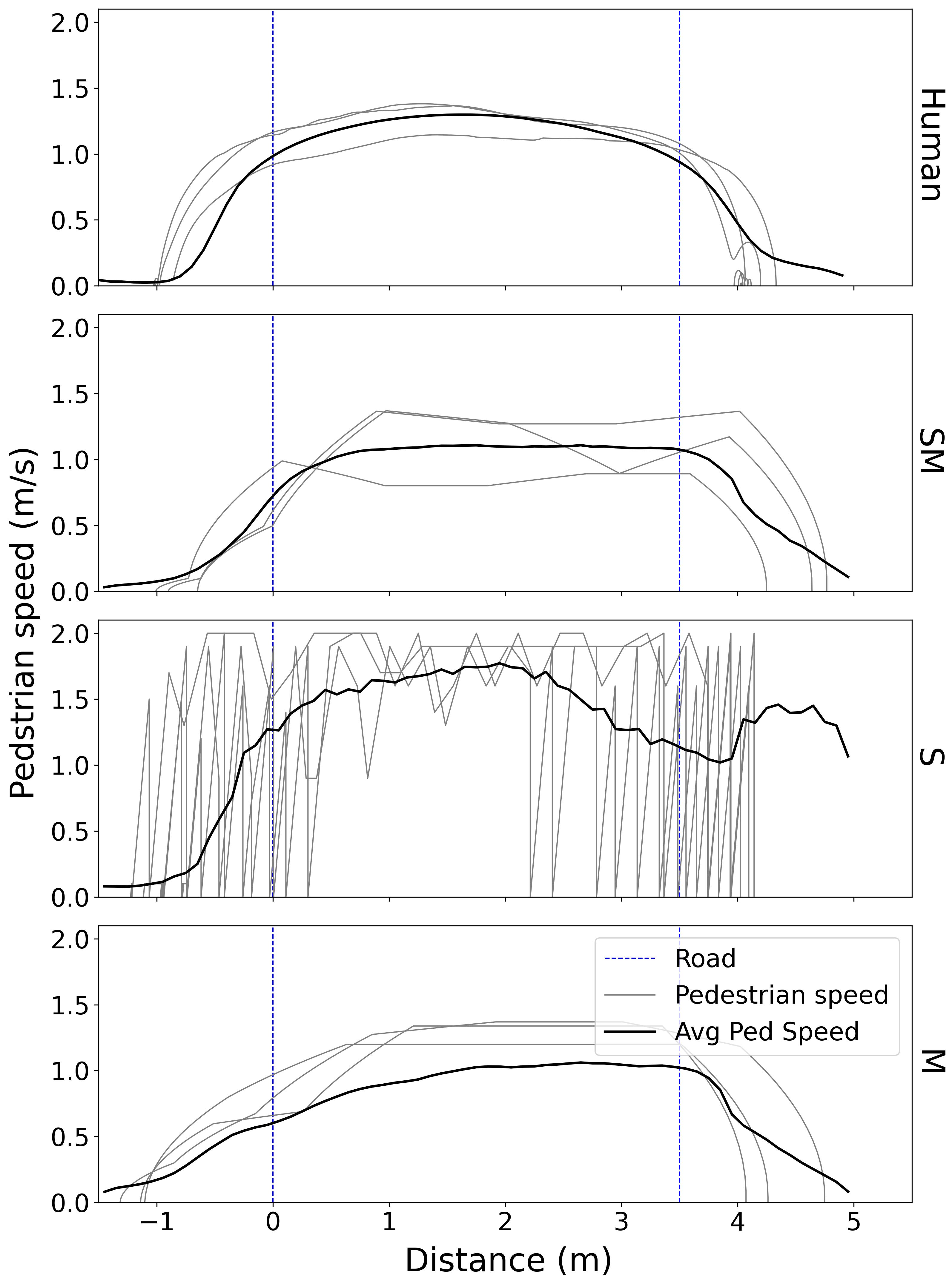}
      \vspace{-0.2cm}
      \caption{Average walking speed profile of the experiment, and different model variants. gray lines are the randomly selected speed curves and black lines are the averaged speed curves of all conditions. The blue dashed lines denote the road curbs, and the walking direction is from left to right.}
      \label{fig: Spd_dis_Combined}
      \vspace{-0.4cm}
\end{figure}

\subsubsection{CIT}

Rows 2-4 of \figurename\ref{fig: CIT_Combined} show the CIT of the three model variants: The SM model exhibited the best performance by accurately capturing four key phenomena. It shows the increase in CIT with both speed and time gaps in non-yielding conditions, aligning closely with human data. In early crossing conditions within yielding scenarios, it captured the speed-dependent pattern, and in late crossing scenarios and illustrated the relationship between CIT and time gaps in late crossing scenarios. The S model captured two phenomena: the increase in CIT with time gaps in non-yielding and early crossing conditions but failed to capture the phenomena observed in late crossing scenarios or the effects of daytime and eHMI. The M model only showed the increase in CIT with vehicle speed.

\subsubsection{Average walking speed}

Rows 2-4 of \figurename\ref{fig: Spd_Combined} show the average walking speed of the three model variants: The SM model captured all three phenomena successfully. The M model captured two phenomena: the pattern where the agent walked slower in late crossing conditions and with larger time gaps. However, the S model did not capture any of the observed phenomena related to average crossing speed.

\subsubsection{The pedestrian speed profile}

\figurename\ref{fig: Spd_dis_Combined} shows the speed profile of the experiment and the three model variants. It can be seen that the integration of walking effort and ballistic speed control in the SM and M models generated smooth, bell-shaped, human-like walking speed curves. These models captured the gradual acceleration and deceleration phases, resulting in a more realistic representation of pedestrian behavior. Whereas, for the S model, shown in the third row of \figurename~\ref{fig: Spd_dis_Combined}, using direct speed control \cite{vizzari2022pedestrian,Sun2019Crowd,xu2020local} rather than the biomechanical model of walking, there were abrupt fluctuations in the agent's walking speed. This is unrealistic for human locomotion, and as a result, the human-like speed profile was not captured by this model.

\section{Discussion}
\subsection{Main findings}

In this study, we developed a pedestrian crossing decision model that integrates RL with sensory-motor mechanisms to simulate human-like road crossing behavior. We identified three main findings:

As shown in \figurename~\ref{fig: Summary_table}, the SM model variant captures most of the phenomena observed in the human data, including the gap acceptance and early crossing behavior, the relation between the CIT and time gap and vehicle speed, and the phenomena related to the average crossing speed. 

Unlike previous models considering the cognitive process, which mostly focused on the binary 'go' or 'not go' decision, our model captures the entire locomotion process of crossing the road, which can be beneficial to pedestrian behavior simulation and prediction.

Building on our previous crossing decision model, which primarily integrated visual limitations with RL, the model proposed here successfully captured several phenomena related to motor aspects of pedestrian locomotion, the day/ night difference and the effect of eHMI, without losing the ability of the simpler model to capture sensory-related phenomena. This is promising, since it shows the extensibility of the overall approach of combining mechanistic modeling of sensory-motor aspects with RL, suggesting that it may be generalisable to more complex traffic scenarios.

\subsection{Impact of sensory-motor constraints on the pedestrian crossing locomotion}

\subsubsection{Human-like walking speed pattern}
Our SM model, with the integration of walking effort and ballistic speed control mechanisms, successfully captured the observed phenomena regarding average walking speed and showed similar walking speed profile. Previous studies argued that the relation between the time gap and the crossing speed was caused by the time pressure. However, when comparing the S model and M model, we found that the model with time pressure but without walking effort and ballistic speed control assumptions could not generate that speed pattern. Therefore, we argue that the time gap-dependent walking speed pattern is the result of both the time pressure and motor aspects, modelled here in terms of walking effort and ballistic speed control. 

In addition, the comparison between the S and M models further highlights the necessity of modeling human motor control to accurately simulate pedestrian locomotion. An interesting question for future work is determining the level of detail needed in the motor components of pedestrian crossing models, given that human locomotion is more complex than we have modeled and that more sophisticated human biomechanical simulations have been modelled by RL \cite{song2021deep}.

\subsubsection{Day-night time difference in gap acceptance and early crossing rate}
The SM model captures the day-night time difference in gap acceptance and early crossing rate by fitting each participant with two visual-related non-policy parameters. This approach allowed us to account for variations in visual perception between day and night scenarios. We initially hypothesised that higher crossing rates during the daytime were due to smaller visual noise. However, we found no clear correlation was present between $\sigma_\mathrm{v}$ value and gap acceptance rate, suggesting that the gap acceptance of the model depends on a non-trivial interaction between $\sigma_\mathrm{v}$ and one or more of the other non-policy parameters.

An interesting aspect of our bounded optimality approach is its interpretability compared to data-driven neural network models. While our method is certainly more interpretable, providing insights into why the RL policy behaves in a near-optimal manner in specific situations, it may not always be immediately obvious why this is the case. Understanding the interaction between various parameters in our model highlights the complexities involved and emphasises importance of further refining our framework to enhance our understanding of pedestrian behavior dynamics, potentially leading to more robust and transparent decision-making models in the future.



\subsubsection{Importance of integrated sensory-motor modeling}

Neither the S model nor the M model captured as many phenomena as the SM model. This suggests that modeling realistic pedestrian crossing patterns requires a comprehensive understanding of the constraints faced by humans. The SM model's success in capturing a wide range of phenomena can be attributed to its incorporation of both sensory and motor mechanisms. Our comparison of models with different assumptions demonstrates that only by fully grasping and modeling these complex sensory-motor interactions can we achieve a more accurate and human-like behavioral model.

\subsection{Implications and limitations}

Our model successfully captures many observed phenomena, which suggests that it may be useful for a number of potential practical applications. It can contribute to the prediction of pedestrian behavior in real-time AV algorithms, enhancing the ability of AVs to interact safely and effectively with pedestrians \cite{camara2020pedestrian}. Furthermore, the model can be integrated into virtual testing environments, providing a valuable tool for evaluating AV systems under various pedestrian crossing scenarios \cite{Rasouli2018AutonomousVT}. This can facilitate the development of more robust and reliable AV technologies, ultimately contributing to safer road environments. Additionally, the model has broader applications in traffic safety modeling. Accurately simulating pedestrian behavior can help the design of safer road infrastructures and traffic management systems, improving overall traffic safety and efficiency \cite{Rasouli2018AutonomousVT,camara2020pedestrian}.

However, not all of the behavioral phenomena observed in the empirical study were captured by our SM model. We calculated the effect size (Cohen's d and Cohen's h) for each statistically significant variable. The effect sizes for the missing behavioral phenomena were relatively small ($< 0.40$), suggesting that these variables do not have a substantial impact on the behavioral metrics. An exception was the effect of the time gap on CIT in early crossing scenarios in yielding conditions, where discrepancies may arise from the bimodal pattern criteria derived from experimental data not aligning well with the model's bimodal patterns. Furthermore, in the empirical study there were also some statistically significant differences between young and old participants \cite{lee2024Under}. Preliminary tests with the SM model indicated that it could not easily capture these effects, possibly because we have not identified the most sensitive sensory-motor mechanism affecting these age groups differently. These limitations highlight areas for future improvement and refinement of the model.

\section{Conclusion}
This research presents an RL model that incorporates sensory-motor mechanisms aimed at simulating human-like pedestrian crossing behavior, which is important for the safe deployment of AVs in urban environments, and also has applications in traffic safety more broadly. The experimental findings of this study reveal that pedestrians’ crossing rates and walking speeds vary in response to time gaps and vehicle speeds, and differ between day and night conditions, as well as with the presence of eHMI. Our model successfully integrated a range of sensory-motor constraints, including visual limitations, looming aversion, time pressure, walking effort, and ballistic speed control, allowing it to replicate the interaction and locomotion patterns observed in the experiment. We find that empirically observed time-gap-dependent walking speed patterns can be understood as arising from a trade-off between time pressure and walking effort, captured by our model. The ability of our model to simulate a large number of observed phenomena highlights the versatility of RL in modeling complex human behaviors. It provides insights into the effect of sensory-motor mechanisms on pedestrian-vehicle interactions. Notably, the model extends the phenomena captured in previous studies, demonstrating its capability to generalise across more complex pedestrian behaviors and scenarios. While the model replicates many observed phenomena, it has limitations, such as capturing the influence of eHMI on CIT, and some CIT patterns in yielding scenarios. Nonetheless, by providing a more accurate representation of pedestrian behavior, this research not only contributes to the field of pedestrian behavior modeling but also has the potential to improve AV algorithms and virtual testing environments, ultimately enhancing the coexistence of AVs and pedestrians in shared spaces.


%





\ifCLASSOPTIONcaptionsoff
  \newpage
\fi



%




\bibliographystyle{IEEEtran}
\bibliography{IEEEexample}
%
\vspace{-0.3cm}
\begin{IEEEbiography}[{\includegraphics[width=1in,height=1.25in,clip,keepaspectratio]{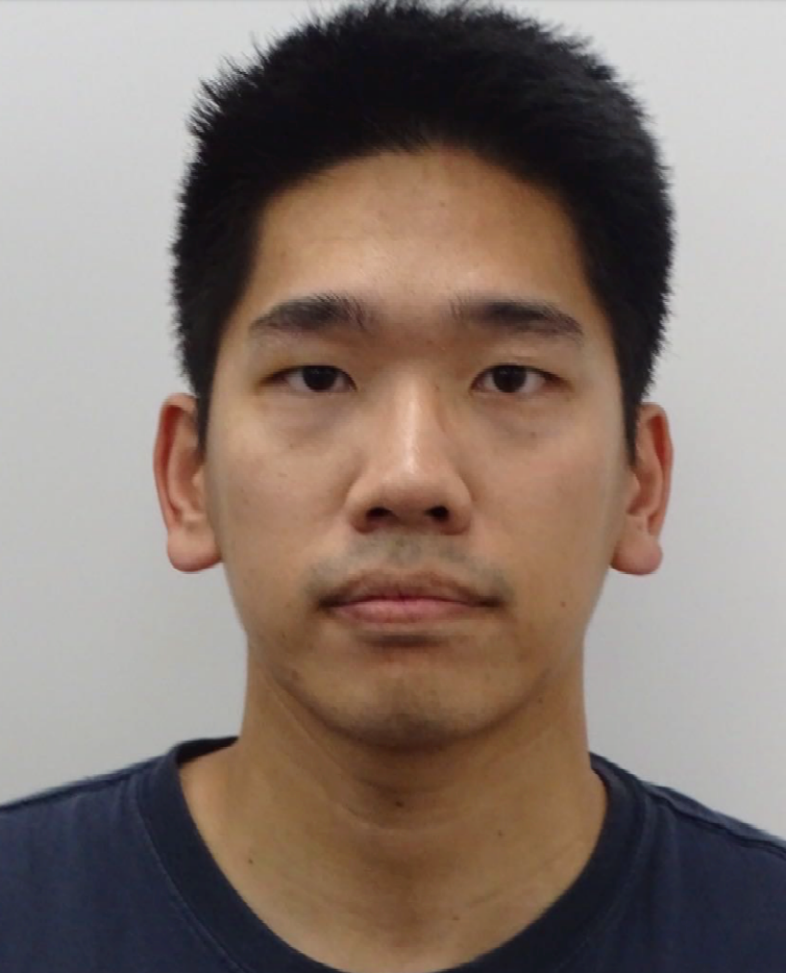}}]
{Yueyang Wang}
received his M.Sc. in Automotive Engineering from the University of Leeds, UK, where he is currently advancing towards a Ph.D. in Transport Studies. His research intersects human factors and safety, with a strong focus on computational modeling of road user behavior and reinforcement learning.
\end{IEEEbiography}

\vspace{-0.3cm}
\begin{IEEEbiography}[{\includegraphics[width=1in,height=1.25in,clip,keepaspectratio]{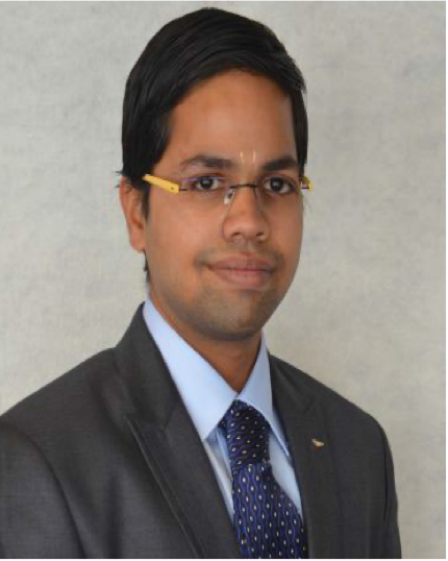}}]
{Aravinda Ramakrishnan Srinivasan}
received the B.Tech. degree in electronics and communication engineering from SASTRA University, Tirumalaisamudram, Tamil Nadu, India, and the M.S. and Ph.D. degrees in mechatronics and mechanical engineering from the University of Tennessee, Knoxville, TN, USA. Before joining the Human Factors and Safety Group, Institute for Transport Studies, University of Leeds, U.K., as a Research Fellow, he was a Post-Doctoral Research Fellow with the Lincoln Centre for Autonomous Systems, University of Lincoln, U.K. His research interests include machine- learning, artificial intelligence, autonomous vehicles, and robotics applications in everyday life.

\end{IEEEbiography}

\vspace{-0.3cm}
\begin{IEEEbiography}[{\includegraphics[width=1in,height=1.25in,clip,keepaspectratio]{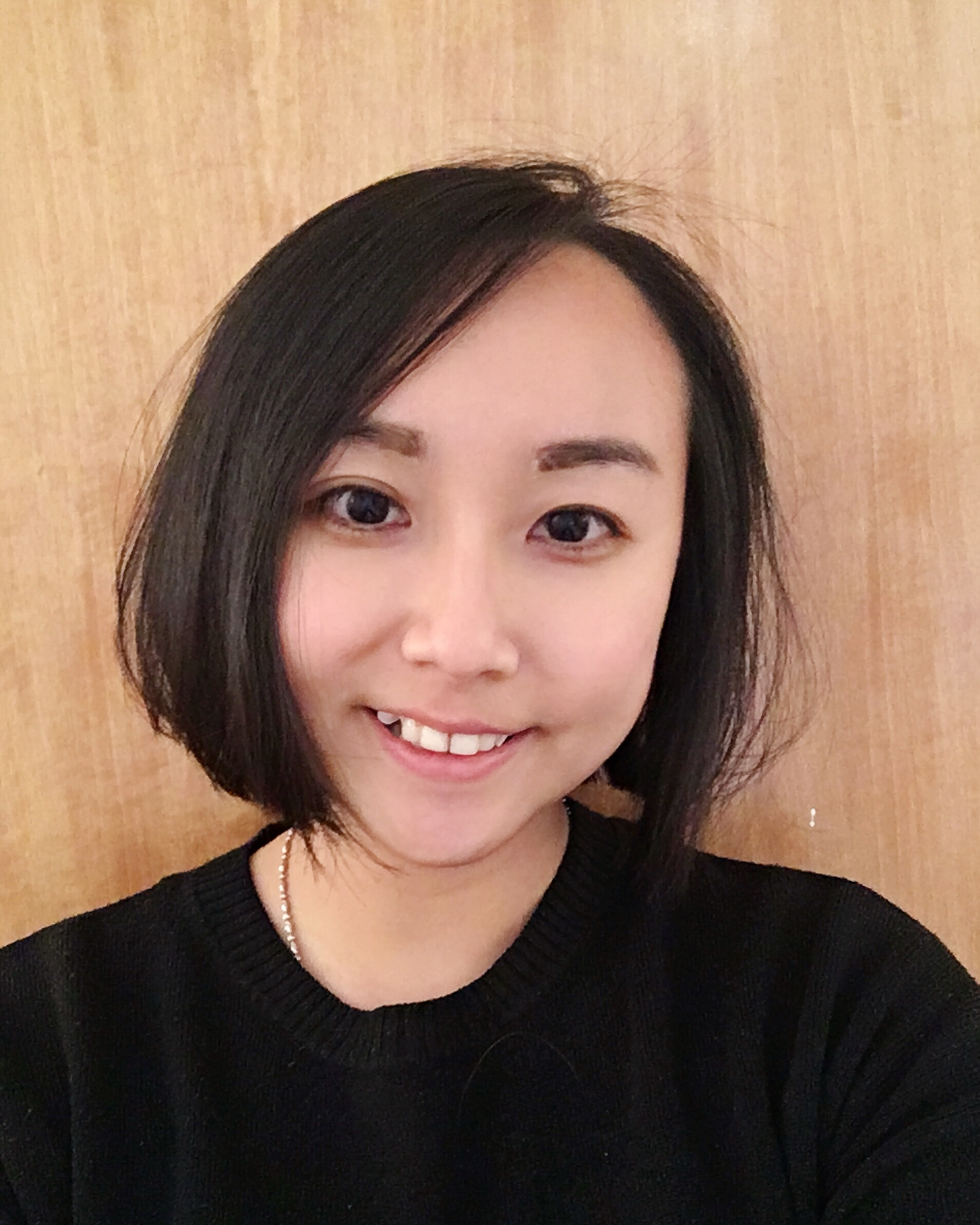}}]
{Yee Mun Lee}
is currently a senior research fellow at the Institute for Transport Studies, University of Leeds. She obtained her BSc (Hons) and her PhD degree from The University of Nottingham Malaysia. Her current research interests include investigating the interaction between automated vehicles and other road users using various methods, especially virtual reality experimental designs. She is involved in multiple EU-funded projects and is actively involved in the International Organisation for Standardisation (ISO).
\end{IEEEbiography}

\vspace{-0.3cm}
\begin{IEEEbiography}[{\includegraphics[width=1in,height=1.25in,clip,keepaspectratio]{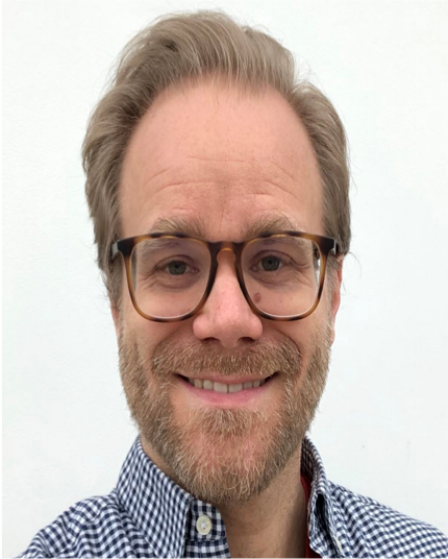}}]
{Gustav Markkula}
received the M.Sc. degree in engineering physics and complex adaptive systems and the Ph.D. degree in machine and vehicle systems from the Chalmers University of Technology, Gothenburg, Sweden, in 2004 and 2015, respectively. He has more than a decade of research and development experience from the automotive industry. He is currently the Chair of Applied Behavior Modeling with the Institute for Transport Studies, University of Leeds, U.K. His current research interests include quantitative modeling of road user behavior and interaction, and virtual testing of vehicle safety and automation technology.
\end{IEEEbiography}







\end{document}